\documentclass[prl,showpacs]{revtex4}
\begin{document}

\title{Nonlinear electromagnetic wave equations for superdense magnetized plasmas}

\author{Nitin Shukla\cite{a)}, G. Brodin, M. Marklund,
P. K. Shukla\cite{b)}, and L. Stenflo\cite{c)}}
\affiliation{Department of Physics, Ume\aa~ University,
SE-90187 Ume\aa,~, SE-90187 Sweden}
\received{2 March 2009}

\begin{abstract}
By using the quantum hydrodynamic and Maxwell equations, we derive nonlinear
electron-magnetohydrodynamic (MHD), Hall-MHD, and dust Hall-MHD equations
for dense quantum magnetoplasmas. The nonlinear equations include the electromagnetic,
the electron pressure gradient, as well as the quantum electron tunneling and electron
spin forces. They are useful for investigating a number of wave phenomena including
linear and nonlinear electromagnetic waves, as well as three-dimensional electromagnetic
wave turbulence spectra arising from the mode coupling processes in dense magnetoplasmas.
\end{abstract}
\pacs{52.25.Dg, 52.27.Gr, 52.35.Mw, 71.10.Ca}

\maketitle

\section{Introduction}

Superdense quantum plasmas are ubiquitous in compact astrophysical objects
\cite{r0,r1,r2,r3} (e.g. the interior of white dwarf stars, magnetars, giant planetary interiors),
and in the next generation intense laser-solid density plasma experiments \cite{r3,r3a,r3b,r4,r5}.
In dense plasmas the degenerate electrons follow Fermi-Dirac statistics,
with quantum tunneling \cite{r6,r7,r8,r8a,r8b,r8c,r9} and spin \cite{r10,r11,r11a,r12,r13}
forces due to the spread in the electron probability wave-function. The quantum statistical
electron pressure and quantum forces produce wave dispersion at nanoscales. Accordingly,
there have been a great deal of interest \cite{r7,r8,r9,r12,r13,r14} in investigating linear
and nonlinear waves and structures at quantum scales in very dense plasmas.

It is well known that superdense plasmas in white dwarf stars and magnetars
are strongly magnetized. Accordingly, the electron dynamics is greatly
affected by the Lorentz force. Furthermore, there is a Bohr magnetization
of electrons which have $1/2$-spin. Thus, new electromagnetic wave theories
(both linear and nonlinear) in dense magnetized plasmas have to be developed,
accounting for the quantum statistical electron pressure, the quantum force
associated with electron tunneling, and the Bohr electron magnetization due
to the spin effect.

In this work, we derive nonlinear equations for electromagnetic waves in superdense
quantum magnetoplasmas. Specifically, we focus on the electron-MHD, Hall-MHD, and dust
Hall-MHD plasmas, and show how the density, fluid velocity, and magnetic fields are
coupled in a non-trivial manner. The present equations are useful for studying numerically
\cite{r15} the linear and nonlinear wave phenomena at quantum scales in superdense
astrophysical plasmas.

\section{Derivation of the nonlinear equations}

The governing nonlinear equations for electromagnetic waves in dense plasmas
are the quantum hydrodynamic equations composed of the continuity equation

$$\frac{\partial n_j}{\partial t} + \nabla \cdot (n_j {\bf u}_j) = 0,
\eqno(1)$$
the electron momentum equation


$$n_e m_e \left[\frac{\partial {\bf u}_e}{\partial t}
+ {\bf u}_e \cdot \nabla {\bf u}_e\right]
= - n_e e \left[{\bf E} + \frac{1}{c}
{\bf u}_e \times {\bf B} \right] - \nabla p_e + {\bf F}_{Qe}
\eqno(2)$$
the Faraday law

$$ c \nabla \times {\bf E} =- \frac{\partial {\bf B}}{\partial t},
\eqno(3)$$
the Maxwell equation including the magnetization spin current

$$
\nabla \times {\bf B} = \frac{4\pi }{c}({\bf J}_p + {\bf J}_m)
+ \frac{1}{c}\frac{\partial {\bf E}}{\partial t},
\eqno(4)
$$
and the momentum equations

$$n_\sigma m_\sigma \left[\frac{\partial {\bf u}_\sigma}{\partial t}
+ {\bf u}_\sigma \cdot \nabla {\bf u}_\sigma \right]
= q_\sigma n_\sigma \left[{\bf E} + \frac{1}{c}
{\bf u}_\sigma \times {\bf B} \right] -n_\sigma m_\sigma \nabla \phi_g,
\eqno(5)$$
where we have introduced the pressure for a non-relativistic degenerate electron gas

$$p_e =\left(\frac{4\pi^2\hbar^ 2}{5m_e}\right)(3/8\pi)^ {2/3}n_e^{5/3},
\eqno(6)$$
and the sum of the quantum tunneling and spin forces

$$
{\bf F}_{Qe} = \nabla \left(\frac{\nabla^ 2\sqrt{n_e}}{\sqrt{n_e}}\right)
-n_e \mu_B \tanh(\xi)\nabla B.
\eqno(7)$$

The gravitational force is

$$\nabla^ 2\phi_g =4 \pi G \sum_{\sigma =i,d}
m_\sigma n_\sigma.
\eqno(8)
$$
where $G$ is the gravitational constant.

In Eqs. (1)--(7), $n_j$ is the number density of the particle species $j$ ($j$
equals $e$ for the electrons, $i$ for the ions, and $d$ for the dust grains),
${\bf u}_e$ is the electron fluid velocity, ${\bf u}_\sigma$ is the fluid velocity
of the species $\sigma$ ($\sigma =i, d$, the index $i$ and $d$ stand for the ions
and charged dust grains), $m_j$ is the mass, $q_\sigma =Z_i e$ for the ions and
$\epsilon Z_d e$ for the dust grains, $Z_i$ is the ion charge state, $e$ is the magnitude of the
electron charge, $Z_d$ is the number of electron charges on a dust grain, $\epsilon = -1 (+1)$ for
negative (positive) dust, $c$ is the speed of light in vacuum, $\mu_B= e\hbar/2m_e$ is the Bohr magneton,
$\hbar$ is the Planck constant divided by $2\pi$, and $B=|{\bf B}|$. We have here introduced the
plasma current density ${\bf J}_p = - e n_e {\bf u}_e + Z_i e n_i {\bf u}_i + \epsilon Z_d e n_d {\bf u}_d$
and the electron magnetization spin current density ${\bf J}_m = \nabla \times {\bf M}$, where the
magnetization for dynamics on a time scale much slower than the spin precession frequency reads
${\bf M} = n_e \mu_B \tanh(\xi) \hat {\bf B}$. Here $\tanh(\xi) =B_{1/2}(\xi)$, $B_{1/2}$ is the
Brillouin function with argument $1/2$ describing electrons of spin $1/2$, $\xi =\mu_B B/k_B T_{Fe}$,
$\hat {\bf B} ={\bf B}/B$, $k_B$ is the Boltzmann constant, and $T_{Fe}$ is the Fermi electron
temperature. We have assumed that the spin orientation has reached the thermodynamical equilibrium
state in response to the magnetic field, which accounts for the $\tanh(\xi)$-factor. On a time
scale shorter than the spin relaxation time scale, the individual electron spins are conserved,
and thus $\tanh(\xi)$ can be taken as constant for an initially inhomogeneous plasma.

Let us first present the generalized nonlinear electron-MHD equations for dense plasmas.
Here the ions and dust grains form the neutralizing background. The wave phenomena
in the EMHD plasma will occur on a time scale much shorter than the ion/dust plasma
and gyroperiods. In equilibrium, we have \cite{r16}

$$n_{e0} = Z_i n_{i0} + \epsilon Z_d n_{d0},
\eqno(8)
$$
where the subscript $0$ stands for the unperturbed value.

The relevant electron-MHD equations are

$$\frac{\partial n_e}{\partial t} + \nabla \cdot (n_e {\bf u}_e) =0,
\eqno(9)
$$
the electron momentum equation (2), Faraday's law (3) and the
electron fluid velocity given by

$${\bf u}_e = \frac{{\bf J}_m}{e n_e}-\frac{c (\nabla \times {\bf B})}{4 \pi e n_e}
+ \frac{\partial {\bf E}}{e n_e \partial t}.
\eqno(10)$$
We observe that the quantum tunneling and spin forces play a role if there are
slight electron density and magnetic field inhomogeneities in dense plasmas.

Second, we derive the modified Hall-MHD equations in the presence of immobile dust grains.
The Hall-MHD equations shall deal with the wave phenomena on a time scale larger than the
electron gyroperiod. The relevant equations are the electron and ion continuity equations,
the inertialess electron momentum equation


$${\bf E} + \frac{1}{c} {\bf u}_e \times {\bf B} + \frac{\nabla p_e}{n_e e}
- \frac{{\bf F}_{Qe}}{n_e e} =0,
\eqno(11)
$$
Faraday's law (3), the ion momentum equation

$$n_i m_i \left[\frac{\partial {\bf u}_i}{\partial t}
+ {\bf u}_i \cdot \nabla {\bf u}_i \right]
= Z_i e n_i \left[{\bf E} + \frac{1}{c}
{\bf u}_i \times {\bf B} \right] -n_i m_i \nabla \phi_g,
\eqno(12)$$
with

$$\nabla^ 2\phi_g = 4\pi G m_i n_i,
\eqno(13)
$$
and
the electron fluid velocity given by

$${\bf u}_e = \frac{Z_i n_i {\bf u}_i}{n_e} + \frac{{\bf J}_m}{e n_e}
-\frac{c (\nabla \times {\bf B})}{4 \pi e n_e},
\eqno(14)
$$
where we have neglected the displacement current since the Hall-MHD plasma deals
with electromagnetic waves whose phase velocity is much smaller than the speed of
light in vacuum.

We now eliminate the electric field from (12) by using (11), obtaining

$$n_i m_i \left[\frac{\partial {\bf u}_i}{\partial t}
+ {\bf u}_i \cdot \nabla {\bf u}_i \right]
= Z_i e n_i \left[\frac{1}{c}
({\bf u}_i-{\bf u}_e) \times {\bf B} - \frac{\nabla p_e}{n_e e}
+ \frac{{\bf F}_{Qe}}{n_e e} \right] -n_i m_i \nabla \phi_g.
\eqno(15)$$

Furthermore, eliminating ${\bf u}_e$ from (15) by using (14), we have

$$n_i m_i \left[\frac{\partial {\bf u}_i}{\partial t}
+ {\bf u}_i \cdot \nabla {\bf u}_i \right]
= Z_i e n_i \left\{\frac{1}{c}
\left[ \frac{\epsilon Z_d n_d}{n_e} {\bf u}_i- \frac{{\bf J}_m}{e n_e} +
\frac{c (\nabla \times {\bf B})}{4\pi e n_e} \right] \times {\bf B}
- \frac{\nabla p_e}{n_e e}
+ \frac{{\bf F}_{Qe}}{n_e e} \right\} - n_i m_i \nabla \phi_g,
\eqno(16)$$
where $n_e = Z_i n_i + \epsilon Z_d n_d$ and $Z_d n_d$ is constant.

Finally, by using (14) we can eliminate ${\bf E}$ from (3), obtaining

$$\frac{\partial {\bf B}}{\partial t} = \nabla \times \left[
\frac{Z_i n_i {\bf u}_i \times {\bf B}}{n_e}
+ \frac{{\bf J}_m \times {\bf B}}{en_e}
-\frac{c}{4\pi e n_e} \left(\nabla \times {\bf B}\right)\times {\bf B} \right].
\eqno(17)
$$

The ion continuity equation, Eqs. (16) and (17), together with (13) and the quasi-neutrality
condition $n_{i1} = n_{e1}$, where $n_{e1,i1} \ll n_{e0,i0}$ are the desired generalized
nonlinear Hall-MHD dense self-gravitating plasma with immobile charged dust grains.
show a nontrivial linear and nonlinear coupling between the density, ion fluid velocity and
magnetic fields fluctuations. They describe the dynamics of a broad range of electromagnetic
waves in dense quantum plasmas.

Third, we consider a dense magnetoplasma composed of inertialess electrons and ions, as well inertial
dust grains. The dust mass density $n_d m_d$ is supposed to be much larger than the ion mass density
$n_i m_i$. The relevant dust Hall-MHD equations, valid for the low-frequency
(in comparison with the ion gyrofrequency) electromagnetic waves, are then composed of the dust
continuity equation


$$\frac{\partial n_d}{\partial t} + \nabla \cdot (n_d {\bf u}_d) = 0,
\eqno(18)$$
the inertialess electron and ion momentum equations, which are combined with
the dust momentum equation (in which the electric field is eliminated by using the
inertialess electron and ion momentum equation)

$$n_d m_d \frac{d{\bf u}_d}{dt}
= \left(\frac{1}{4\pi} \nabla \times {\bf B}
- \frac{1}{c} {\bf J}_m\right) \times {\bf B}
- \frac{\epsilon Z_d n_d \nabla p_e}{n_e}
+ \frac{\epsilon Z_d n_d {\bf F}_{Qe}}{n_e}
-n_d m_d \nabla \phi_g,
\eqno(19)$$
with

$$\nabla^ 2\phi_g \approx 4\pi G n_d m_d,
\eqno(20)
$$
together with Faraday's law (3). Here $d/dt =(\partial/\partial t) + {\bf u}_d \cdot \nabla$.
In deriving (19) we have used the modified Amp\`ere's law [e.g. Eq.(4) without the displacement 
current] in view of the low-phase velocity (in comparison with $c$) electromagnetic waves.

From Faraday's law (3) and the dust momentum equation, we obtain


$$\frac{\partial {\bf B}}{\partial t}
= \nabla \times \left({\bf u}_d \times {\bf B}\right)
-\frac{c m_d}{q_d}\nabla \times
\frac{d{\bf u}_d}{dt}.
\eqno(21)
$$
Equations (18), (19) and (21) with $n_e \approx Z_d n_d \gg Z_i n_{i0}$
are the desired dust Hall-MHD equations in dense magnetized plasmas.
These equations govern the dynamics of the modified coupled dust-cyclotron,
dust acoustic and dispersive dust Alfv\'en waves. The dispersion arises
due to the quantum electron tunneling and the finite frequency (in comparison
with the dust gyrofrequency) or the dust skin effect.

\section{summary}

In this paper, we have derived the nonlinear equations for electromagnetic waves
in dense self-gravitating magnetoplasmas. For this purpose, we have used the
generalized quantum hydrodynamical and Maxwell's equations, and obtained three sets of 
nonlinear equations, which exhibit the non-trivial linear and nonlinear couplings between
the plasma number density, fluid velocity, and magnetic field fluctuations. The present set of
nonlinear equations should be used to investigate numerically the dynamics of
obliquely (against the external magnetic field direction) propagating modified
electron whistlers with the generalized electron-MHD plasma model, as well as the modified fast
and slow modes and dispersive electromagnetic ion-cyclotron-kinetic Alfv\'en waves \cite{r17,r18} 
within the generalized Hall-MHD plasma model, and the modified coupled dust-cyclotron-dust acoustic
and dispersive dust Alfv\'en waves within the dust Hall-MHD plasma model. The obtained results would 
then provide valuable information on multi-dimensional electromagnetic fluctuation spectra at
nanoscales, which may appear in the interior of white dwarf stars and in magnetars.

\acknowledgements
This work was partially supported by the Deutsche Forschungsgemeinschaft (Bonn)
through the project SH21/3-1 of the Forschergruppe FOR 1048, as well as by the European
Research Council under Contract No. 204059-QPQV, and by the Swedish Research Council under
Contracts No. 2005-4967 and No. 2007-4422.

\end{document}